\documentclass[aps,pra,amsmath,amsfonts,amssymb,longbibliography,reprint,10pt]{revtex4-1}

\usepackage[english]{babel}
\usepackage[utf8]{inputenc}
\usepackage{graphicx}
\usepackage[colorinlistoftodos]{todonotes}
\usepackage{tensor}
\usepackage{epstopdf}
\usepackage{enumerate}
\newcommand{\op}[1]{\hat{#1}}                                 
\newcommand{\ket}[1]{\lvert #1\rangle}                        
\newcommand{\bra}[1]{\langle #1 \rvert}                       
\newcommand{\braket}[2]{\langle #1 \vert #2 \rangle}          
\newcommand{\abs}[1]{\left\lvert #1 \right\rvert}             
\newcommand{\pr}[1]{\ket{#1}\bra{#1}}                         
\newcommand{\mean}[1]{\langle #1 \rangle}                     
\newcommand{\Tr}[1]{\text{Tr}(#1)}                            

\begin{document}
\title{Weak Values are Interference Phenomena}
\date{\today}
\author{Justin Dressel}
\affiliation{Department of Electrical and Computer Engineering; University of California, Riverside, CA 92521, USA.}

\begin{abstract}
Weak values arise experimentally as conditioned averages of weak (noisy) observable measurements that minimally disturb an initial quantum state, and also as dynamical variables for reduced quantum state evolution even in the absence of measurement.  These averages can exceed the eigenvalue range of the observable ostensibly being estimated, which has prompted considerable debate regarding their interpretation.  Classical conditioned averages of noisy signals only show such anomalies if the quantity being measured is also disturbed prior to conditioning.  This fact has recently been rediscovered, along with the question whether anomalous weak values are merely classical disturbance effects.  Here we carefully review the role of the weak value as both a conditioned observable estimation and a dynamical variable, and clarify why classical disturbance models will be insufficient to explain the weak value unless they can also simulate other quantum interference phenomena.
\end{abstract}

\maketitle


\section{Introduction}
After their introduction over a quarter-century ago \cite{Aharonov1988}, quantum weak values \cite{Dressel2014,Kofman2012} have consistently found themselves at the center of controversy \cite{Aharonov2008,Aharonov2010}. Indeed, the original paper \cite{Aharonov1988} details how one can postselect a weak (i.e., noisy) measurement of a spin-$1/2$ operator for an electron (using a sequence of two Stern-Gerlach apparatuses) to obtain a \textit{conditioned expectation value} that approximates a weak value with an anomalously large value of $100$. The question whether this strange average value has any physical meaning pertaining to the spin has since plagued the concept of the weak value (e.g., \cite{Duck1989}). 

The most recent addition to this controversy \cite{Ferrie2014b} considers a superficially similar example consisting of a classical coin that has its two faces noisily measured, then \textit{disturbed}, and finally conditioned to produce an anomalous average value of $100$ heads. The conclusion drawn from their study (which has been heavily criticized \cite{Ferrie2014bcom1,Ferrie2014bcom2,Ferrie2014bcom3,Ferrie2014bcom4,Ferrie2014bcom5}) is that strange weak values may be understood entirely as classical disturbance effects, making them not ``quantum.'' In fact, every element of this simple example of how intermediate disturbance can cause strange postselected averages of noisy signals has been previously demonstrated, and corroborates our published work: Not only did we emphasize a similar disturbance example using a colored marble in our systematic investigation of generalized observable measurements \cite{Dressel2010,Dressel2012b}, but we also carefully highlighted the potential role of invasive measurements in studies linking strange conditioned averages (including weak values) to violations of generalized Leggett-Garg inequalities \cite{Williams2008,Goggin2011,Dressel2011,Groen2013,Dressel2014c} (which were designed to test for ``quantum'' behavior in macroscropic systems \cite{Leggett1985,Leggett2002,Emary2014}). It is now well-established that any hidden-variable model that can produce strange conditioned averages like the weak value must include some form of intermediate disturbance (see also \cite{Tollaksen2007,Ipsen2014}). The more interesting question to raise is not whether a particular strange conditioned average may be explained as classical disturbance, but rather whether such models of disturbance can also reproduce the complete behavior of the weak value as its physical parameters are varied.

In this paper, we revisit this question in order to dispel the abundant confusion about weak values still present in the literature, and emphasize that a strange weak value is nonclassical in precisely the same manner that a single quantum particle can be considered to be nonclassical. Specifically, strange weak values fundamentally arise from \emph{interference} (i.e., superoscillations \cite{Aharonov2011,Berry2012}), and thus also appear in any wave-like field theory, such as classical optics \cite{Ritchie1991,Kocsis2011,Lundeen2011}, in a straightforward way. In such a classical field theory, anomalous weak values do faithfully indicate physical wave properties, despite how counter-intuitive their predictions may seem. For example, the orbital part of the Poynting vector field of optical vortex beams, or evanescent fields, can show anomalous local momentum distributions that are precisely equal to strange weak values \cite{Bliokh2013a,Bliokh2014,Dressel2014b}. Therefore, as with any quantum interference effect, only the fact that discrete and independent random events can be measured (as opposed to attenuated wave intensities) will distinguish whether the statistics producing a strange weak value are truly quantum mechanical in origin.  (Alternatively, entangling the degrees of freedom of distinct particles will not have a simple classical field interpretation, e.g., \cite{Dressel2011}.) Nevertheless, even in the case of discrete measurement events the large number of measurements needed to statistically resolve such a weak value still imply that it is best considered as a dynamical physical variable for the effective (classical) mean field, and not necessarily to each individual quantum particle \emph{a priori} \cite{Dressel2014b}. To emphasize these subtle points, we carefully review several complementary approaches to deriving and understanding the weak value, paying special attention to its role as an ideal estimate, an experimentally measurable conditioned average, and as a classical dynamical variable for reduced quantum state evolution. This detailed treatment aims to supplement the simplified introduction to the experimental applications of weak values in \cite{Dressel2014} with an expanded theoretical discussion that highlights their pervasive and under-appreciated role throughout the quantum formalism.

In what follows we also emphasize the often overlooked connection between weak values and joint quasiprobability distributions (such as the Kirkwood-Dirac \cite{Kirkwood1933,Dirac1945,Chaturvedi2006,Lundeen2012,Hofmann2012b,Lundeen2014}, Terletsky-Margenau-Hill \cite{Terletsky1937,Margenau1961,Johansen2004,Johansen2004c}, and the various Moyal phase space distributions \cite{Wigner1932,Moyal1949,Glauber1963,Sudarshan1963}) that determine conditioned observable estimates. Notably, to obtain a strange weak value outside the usual eigenvalue bounds, these joint quasiprobability distributions must become negative as a consequence of the nonclassical quantum interference between probability amplitudes. The best known examples of this intrinsic negativity from quantum interference occur in the Moyal phase space distributions, such as the Wigner distribution for quadratures, or the Glauber-Sudarshan $P$ distribution for coherent state amplitudes, which show such negativity with nonclassical optical states \cite{Wigner1932,Glauber1963,Sudarshan1963,Moyal1949}. Indeed, for single particles such negativity in quasiprobability distributions has been proven to be an equivalent notion of ``nonclassicality'' as the need for \textit{contextual} hidden variables \cite{Spekkens2008,Ferrie2011} (in the sense of Bell-Kochen-Specker \cite{Kochen1967,Mermin1993}), and formally arises from the usual operator non-commutativity of quantum mechanics. This connection between contextuality and strange weak values was emphasized in a recent proof by Pusey \cite{Pusey2014}, as well as an earlier study by Tollaksen \cite{Tollaksen2007}, and is consistent with the established understanding that strange weak values arise fundamentally from (quantum) interference. It follows that if a classical model as in \cite{Ferrie2014b} could really mimic the detailed functional structure of the weak value, then it would also be able to simulate other features that are normally considered to be quantum mechanical. We further emphasize this latter point by reviewing the deep connections between weak values and the classical dynamical variables of the Hamilton-Jacobi formalism, both in its fully quantum generalization that is equivalent to the Schr\"odinger equation, and in the resulting classical limit.

This paper is organized as follows. In Section~\ref{sec:estimates} we derive and discuss how the weak value is the best statistical estimate for the average (but unmeasured) observable value for the times between two known measurement events. In Section~\ref{sec:measurement} we discuss three approaches for experimentally verifying the weak value as the appropriate such estimate: weak von Neumann coupling, weak generalized observable measurements, and as the physical dynamical variables for reduced state evolution. In Section~\ref{sec:quasiprob} we explicitly connect the weak value to quasiprobabilities, focusing on the Terletsky-Margenau-Hill, Kirkwood-Dirac, and Wigner distributions, and explicitly connect weak values to classical mean-field dynamical variables using the Hamilton-Jacobi quantum-classical correspondence.  We conclude in Section~\ref{sec:conclusion}. 

\begin{figure}[t]
  \includegraphics[width=0.8\columnwidth]{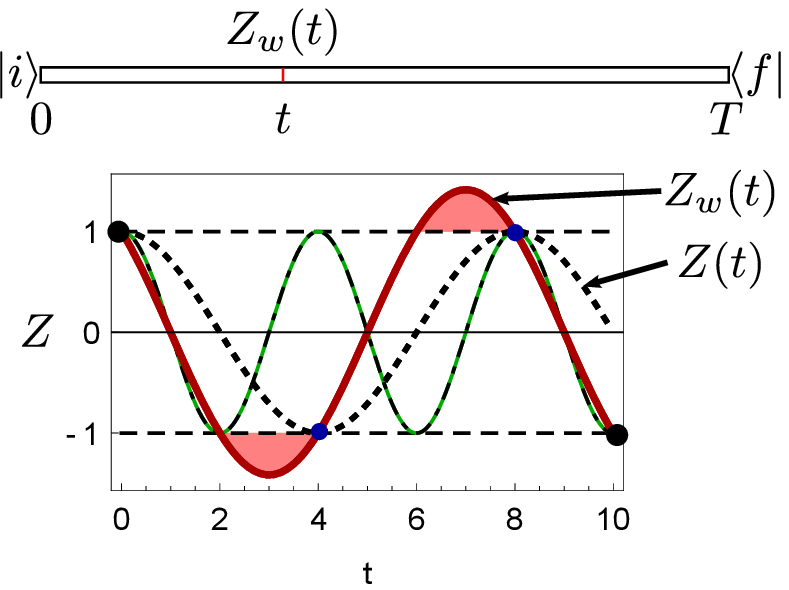}
  \caption{The weak value $Z_w(t) = \text{Re}\bra{f}\op{U}_{T-t}\op{Z}\op{U}_t\ket{i}/\bra{f}\op{U}_T\ket{i}$ estimates $\op{Z}$ conditioned on two boundaries that bracket the time interval $[0,T]$. For the Pauli $\op{Z} = \pr{1}-\pr{0}$ qubit operator prepared in $\ket{i} = \ket{1}$, postselected in $\bra{f} = \bra{0}$, and evolving with $\op{U}_t = \exp[i \omega t (\ket{1}\bra{0}+\ket{0}\bra{1})]$, the weak value $Z_w(t)$ (thin, green) coincides with the expectation value $Z(t) = \bra{i}\op{U}_t^\dagger\op{Z}\op{U}_t\ket{i}$ (dot-dashed, black) when the postselection is consistent with the natural oscillation.  Otherwise, $Z(t)$ (dashed, black) displays a jump at time $T$, while $Z_w(t)$ (red) smoothly connects the boundaries, still passing through the same points of certain $Z$ (blue dots).  The shaded regions exceed the eigenvalue bounds of $\pm 1$, indicating the inconsistency between the natural evolution and the observed boundaries.}
  \label{fig:disturb}
\end{figure}

\section{Weak values as estimates}\label{sec:estimates}
Most of the controversy surrounding weak values rests upon their common (but unnecessary) association with an alternative time-symmetric approach to the quantum theory that involves two state vectors \cite{Watanabe1955,Aharonov1964,Aharonov1990,Aharonov1991,Aharonov2005,Aharonov2009}. In this time-symmetric approach, one forward-propagates a state-vector $\ket{i}$ from an initial time $0$ to $t$ in the usual way; however, one also \textit{back-propagates} a second state-vector $\bra{f}$ from a final time $T$ to $t$. While the initial state vector $\ket{i}$ corresponds to a preparation procedure, the final state vector $\bra{f}$ corresponds to a \textit{postselection} procedure. 

Interestingly, the best estimate \cite{Hall2001,Johansen2004b,Hall2004} of the average (unmeasured) value for an observable $\op{A}$ at any time $t$ in the interval $[0,T]$ is then not the expectation value $A(t) = \bra{i}\op{U}_t^\dagger \op{A} \op{U}_t\ket{i}$ (which neglects the information about the postselection), but is rather the \textit{weak value} 
\begin{align}\label{eq:wv}
  A_w(t) &= \text{Re}\,\frac{\bra{f}\op{U}_{T-t}\op{A}\op{U}_t\ket{i}}{\bra{f}\op{U}_T\ket{i}},
\end{align}
as we will derive shortly. Here $\op{U}_t = \exp(-i\op{H} t/\hbar)$ is the unitary propagator for a time-interval $t$ that is generated by the Hamiltonian $\op{H}$. Note that the imaginary part of the weak value, while independently interesting as a measure of intrinsic measurement disturbance \cite{Johansen2004b,Dressel2012d,Hofmann2014}, is unrelated to the estimation of $\op{A}$ as an observable, so we will ignore it for now.

The problematic feature of Eq.~\eqref{eq:wv} as an estimate is that it may exceed the eigenvalue range of $\op{A}$; such strange behavior is illustrated as the shaded areas in Figure~\ref{fig:disturb}. As discussed in the introduction, a classical conditioned estimate may show such anomalous behavior only if the estimation procedure is \textit{noisy} and if what is being estimated is \textit{disturbed} in the interval $[0,T]$ \cite{Dressel2012b,Tollaksen2007,Williams2008,Dressel2011,Ipsen2014}. The question raised in Ref.~\cite{Ferrie2014b} is whether such a classical model with noisy estimation and disturbance is sufficient to explain Eq.~\eqref{eq:wv}. As we will show in what follows, such a classical disturbance explanation is difficult to defend when the many roles of this weak value expression are examined in detail.

\subsection{Derivation of the best estimate}
For completeness we now show how this weak value formula can be obtained as the optimal estimate that minimizes a statistical uncertainty metric, justifying its interpretation as a best estimate.

Suppose one wishes to estimate the best average values for an observable $\op{A}$ given a initial preparation $\ket{i}$, followed by a projective measurement in a particular basis $\ket{f}$ that does not commute with $\op{A}$.  For each $f$, we can guess a value $\bar{a}_f$ that estimates the average of $\op{A}$ given that the specific result $f$ was observed. This procedure formally constructs a Hermitian observable $\op{A}_{\text{est}} = \sum_f \bar{a}_f\,\pr{f}$ that is measured by the procedure, which contains the estimates for every $f$. The goal is then to determine the optimal such estimates $\bar{a}^{\text{opt}}_f$ of $\op{A}$ conditioned on each measured result $f$.

To accomplish this goal, we must define a measure for how close the estimate $\op{A}_{\text{est}}$ is to the target observable $\op{A}$.  A natural choice for such a measure is the weighted trace distance between two observables \cite{Hall2001,Johansen2004,Hall2004}
\begin{align}
  \mathcal{D}_\rho(\op{A},\op{B}) &= \Tr{\op{\rho}\,(\op{A}-\op{B})^2},
\end{align}
which can generally depend upon any positive prior bias $\op{\rho}$.  We can interpret this measure as specifying the shortest \emph{geometric distance} in operator space, weighted by the \emph{statistical prior} bias of $\op{\rho}$.  Since $\op{A}$ and $\op{B}$ need not commute, there is generally no straightforward operational interpretation for this distance \cite{Busch2013,Dressel2014d,Busch2014}, but it does formally provide a reasonable definition of the geometric ``closeness'' for the two operators.

Now suppose we have a definite prior state $\op{\rho} = \pr{i}$ and choose $\op{B} = \op{A}_{\text{est}}$.  Computing the weighted trace distance yields \cite{Hall2001,Johansen2004,Hall2004}
\begin{align}
  \mathcal{D}_i(\op{A},\op{A}_{\text{est}}) &= \bra{i}\left[\op{A}^2 + \op{A}_{\text{est}}^2 - (\op{A}\op{A}_{\text{est}} + \op{A}_{\text{est}}\op{A})\right]\ket{i}, \nonumber \\
  &= \bra{i}\op{A}^2\ket{i} - \textstyle{\sum_f} \abs{\braket{f}{i}}^2\,\left[\text{Re}\frac{\bra{f}\op{A}\ket{i}}{\braket{f}{i}}\right]^2 \nonumber \\
  &\quad + \textstyle{\sum_f}\abs{\braket{f}{i}}^2\,\left[\bar{a}_f - \text{Re}\frac{\bra{f}\op{A}\ket{i}}{\braket{f}{i}}\right]^2. 
\end{align}
Only the final term depends on the choice of estimates $\bar{a}_f$ for each $f$, and is positive definite. Therefore, the trace distance is minimized when this term vanishes, which in turn implies that the optimal estimate for each independent $f$ must be the weak value formula
\begin{align}
  \bar{a}^{\text{opt}}_f &= \text{Re}\frac{\bra{f}\op{A}\ket{i}}{\braket{f}{i}}.
\end{align}
Note that if the basis $\ket{f}$ is chosen to be the eigenbasis of $\op{A}$, then these optimal estimates reduce identically to the eigenvalues of $\op{A}$ and the trace distance vanishes.

To obtain Eq.~\eqref{eq:wv} including intermediate time-evolution, we can choose a particular $\ket{f} \mapsto \op{U}_{T-t}\ket{f}$ and initial state $\ket{i} \mapsto \op{U}_t\ket{i}$ that take into account Hamiltonian propagation $\op{U}_t = \exp(t\op{H}/i\hbar)$ from the observed results for a specific preparation and postselection. 

This pure-state derivation may also be generalized in a straightforward way \cite{Hall2004}, which produces a formulation of the weak value suitable for generalized measurements and mixed states
\begin{align}\label{eq:wvgen}
  A_w(t) &= \text{Re}\,\frac{\Tr{\op{E}_{T-t}\,\op{A}\,\op{\rho}_t}}{\Tr{\op{E}_{T-t}\,\op{\rho}_t}}.
\end{align}
Here the back-propagating operator $\op{E}_{T-t}$ is often called a ``retrodictive state,'' or ``effect matrix,'' in contrast to the forward-propagating ``predictive state'' $\op{\rho}_t$, or ``density matrix.'' Recently, the estimate in Eq.~\eqref{eq:wvgen} has been used to great effect experimentally \cite{Campagne-Ibarcq2014,Weber2014,Tan2014,Rybarczyk2014} for ``quantum smoothing'' \cite{Tsang2009,Tsang2012} and ``past quantum state'' analyses \cite{Gammelmark2013,Gammelmark2014} of continuously measured signals (e.g., it was used to track individual photon emissions into a monitored cavity \cite{Rybarczyk2014}). Both $\op{E}_{T-t}$ and $\op{\rho}_t$ generally evolve according to open-system master equations \cite{Breuer2007,Wiseman2009} that can also include the effects from additional (discrete or continuous-in-time) stochastic measurement-results \cite{Dressel2013b,Gammelmark2013,Chantasri2013}, in contrast to the closed-system (unitary) Schr\"odinger-von Neumann dynamics usually assumed with Eq.~\eqref{eq:wv}.  Note that if the effect matrix $\op{E}_{T-t}$ is the identity $\op{1}$, then no posterior conditioning has been performed, so the usual expectation value is also recovered as a special case.  

\subsection{Interpreting and generalizing the estimate}
As a philosophical side note, for those who believe that the state-vector represents the complete physical (ontic) reality (e.g., adherents to the many-worlds interpretation \cite{Vaidman2014}), this time-symmetric estimate prompts several more radical speculations: The existence of the second state vector $\bra{f}\op{U}_{T-t}$ in Eq.~\eqref{eq:wv} seems to imply not only that the state $\op{U}_t\ket{i}$ is an incomplete description of reality at time $t$, but also that there seems to be a causal effect on the time $t$ from the future time $T$ \cite{Aharonov2012}. Such a \emph{retro-causal} interpretation is similar in spirit to the interpretations of anti-particles in quantum field theory as field-excitations that move backwards through time \cite{Feynman1949}.  However, just as with anti-particles, one does not need to invoke such controversial philosophical concepts as physical state-vectors or retro-causation to meaningfully interpret the weak value in Eq.~\eqref{eq:wv} as the best available statistical estimate given only the information about the specified boundary conditions.

A more pragmatic attitude (which we shall adopt here) is to treat the estimate in Eq.~\eqref{eq:wv} as \textit{subjective} (epistemic), and pertaining to a time interval $[0,T]$ that has already occurred in the past.  That is, one performs an experiment that prepares $\ket{i}$ at time $0$, waits a duration $T$, then makes a projective measurement that shows a result corresponding to the state $\bra{f}$. One then interprets Eq.~\eqref{eq:wv} as the best estimate of the (unmeasured) average value of $\op{A}$ within that time interval \cite{Hall2001,Johansen2004b,Hall2004}, given only the knowledge of $\ket{i}$, $\bra{f}$, and $\op{H}$. We emphasize that this approach is no different in character than stating that the expectation value $\bra{i}\op{U}^\dagger_t\op{A}\op{U}_t\ket{i}$ is the best estimate for the (unmeasured) average value of $\op{A}$, given only the knowledge of the preparation $\ket{i}$ and $\op{H}$.  Indeed, such a counterfactual interpretation of the expectation value in the absence of measurement is at the core of the Ehrenfest theorem that equates quantum expectation values with mean-field classical dynamical variables \cite{Ehrenfest1927}.  If we interpret weak values as similar (but additionally constrained) classical dynamical variables, then an anomalous weak value should indicate the presence of some interesting intermediate physical process that must have occurred in order to satisfy both boundary conditions that bracket the time interval $[0,T]$ (see Figure~\ref{fig:disturb}).

Supporting this point of view is the fact that similar bidirectional (in time) estimates about unknown properties of structured stochastic processes (e.g., hidden Markov models) during such an interval are now well-established in classical computational mechanics \cite{Shalizi2001,Crutchfield2009,Ellison2009,Mahoney2011}. There it is shown that one should use both forward and reverse ``causal states'' (i.e., probability distributions) that contain information gathered both before \textit{and after} each time $t$ to optimally estimate the properties of an evolving stochastic process. Similarly, classical statistics and filtering theory also use bidirectional states to provide the best estimate for information contained in noisy data confined to a time interval (called optimally ``smoothing'' the noise) \cite{Simonoff1998,Einicke2012}. Since quantum theory is closely related to probability theory \cite{Dressel2012b,Leifer2013}, it is logical that similar estimation methods can be applied.  Indeed, upgrading these estimation schemes to the quantum realm \cite{Pegg2002a,Dressel2013b,Leifer2013} produces both states in Eq.~\eqref{eq:wv}, as well as the mixed-state generalization of Eq.~\eqref{eq:wvgen} \cite{Pegg2002b,Wiseman2002,Coecke2012}.

\section{Measuring weak values}\label{sec:measurement}
The confidence that estimations like the expectation value, or the weak value in Eqs.~\eqref{eq:wv} and \eqref{eq:wvgen}, reflect something meaningful about the physical world (and are not merely fevered hallucinations of the mind) follows from verification of their predictions by experimental measurements. In the case of the expectation value, any unbiased estimation of $\op{A}$ will suffice, corroborating the predicted result. In the case of the weak value, however, the presence of the posterior boundary condition additionally constrains the form of the possible measurements that can verify the estimate. 

Specifically, those measurements must be ``weak,'' meaning that they should not appreciably perturb the evolution of the quantum system. Since information extraction necessarily disturbs the quantum state, only minimally informative (i.e., \emph{noisy}) measurements will leave the state mostly unperturbed \cite{Aharonov1988,Wiseman2009}, and thus faithfully reproduce the assumptions made about the evolution during the time interval $[0,T]$ by the formulas in Eqs.~\eqref{eq:wv} and \eqref{eq:wvgen}. The surprising fact is that averaging such weak observable measurements can indeed consistently verify the weak value as the correctly estimated average, even when it predicts anomalous averages.

In what follows we will detail the standard von Neumann approach for measuring the weak value, as well as a more general approach that solidifies its interpretation as a conditioned average in the limit of negligible disturbance to the quantum state.  We will also detail how the weak value appears as a classical dynamical variable for reduced system evolution even outside the usual context of postselected weak measurements, further cementing its interpretation as the appropriate mean-field variable that physically estimates the observable when the natural state evolution is unchanged by external influences.

\subsection{von Neumann interaction}
The standard approach for performing a weak observable measurement \cite{Aharonov1988,Dressel2012e,Kofman2012,Dressel2014}, originally due to von Neumann \cite{vonNeumann1932}, is to couple the \emph{system} observable of interest $\op{A}$ (such as the spin of a particle) to a \emph{detector} observable $\op{F}$ (such as the transverse momentum of the same particle) for an independent degree of freedom, using a simple linear interaction Hamiltonian
\begin{align}\label{eq:vnham}
  \op{H}_{DS}(t) &= \hbar\, g(t)\,\op{F}\otimes\op{A}.
\end{align}
The time-dependent coupling profile $g(t)$ is typically assumed to be zero outside a short interval of duration $\delta t$ and to be impulsive, i.e., short on the timescale of the natural dynamics of both the system and the detector.  In the interaction picture for the system and detector, this coupling Hamiltonian produces a joint unitary rotation of the joint state that entangles the system with the detector
\begin{align}\label{eq:interaction}
  \op{V}_{DS} &= \exp(-ig\,\op{F}\otimes\op{A}),
\end{align}
where $g = \int_0^{\delta t}\!\! g(t')\,dt'$ is the effective coupling strength for the impulsive interaction.  

For a concrete example of such an interaction, in the optical experiments \cite{Ritchie1991,Kocsis2011,Lundeen2011} a wafer of birefringent crystal was used to couple the polarization (system) of a paraxial beam to its continuous transverse momentum (detector).  In contrast, the optical experiments \cite{Pryde2005,Goggin2011,Dressel2011} used polarization-dependent reflection from partially transmitting optical elements to couple the polarization (system) of a paraxial beam to its binary orbital which-path degree of freedom (detector).  In the more recent superconducting qubit experiment \cite{Groen2013} the energy basis of one transmon qubit (system) was coupled via an intermediate bus stripline resonator to the binary energy basis of a second and physically separated transmon qubit (detector) using microwave pulses.

\begin{figure}[t]
  \includegraphics[width=\columnwidth]{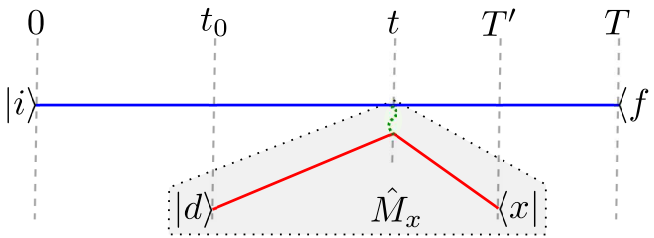}
  \caption{Impulsive von Neumann interaction for measuring $\op{A} = \sum_a a\,\pr{a}$.  A system degree of freedom $\ket{i}$ (blue, top) is prepared at time $0$, and a detector degree of freedom $\ket{d}$ (red, bottom) is prepared at a potentially different time $t_0$.  After independent unitary propagation $\op{U}_t$ and $\op{U}_{t-t_0}^{(D)}$, respectively, to the intermediate time $t$ they interact impulsively (green, wavy) with a joint unitary interaction $\op{V}_{DS}$ for a short duration $\delta t$.  After the interaction, both system and detector continue to propagate until the detector is measured at a time $T'$ to find the result $\bra{x}$, and the system is measured at a potentially different time $T$ to find the result $\bra{f}$. This entire coupling procedure may be represented as a single system operation (lightly shaded region) $\op{M}_x = \bra{x}\op{U}^{(D)}_{T'-t}\op{V}_{DS}\op{U}^{(D)}_{t-t_0}\ket{d}$ that includes the detector preparation, evolution, coupling, and measurement together, producing a net effect only on the evolving system state. If the identity $\op{A} = \sum_x \alpha_x\,\op{M}^\dagger_x\op{M}_x$ can be satisfied for some calibrated signal values $\alpha_x$ for the detector, then the system observable $\op{A}$ can be faithfully estimated by the detector results $x$ in an unbiased way. For strong measurements by the detector, the effective system operations become eigenstate projections $\op{M}_x \to \pr{a}$ with an eigenvalued signal $\alpha_x \to a$, while for weak measurements the operations approximate the identity $\op{M}_x \approx \op{1}$ with a noisy signal $|\alpha_x|\gg|a|$, leaving the state essentially unperturbed. Averaging this noisy signal $\alpha_x$ for such weak measurements conditioned on a particular $f$ approximates the weak value $\text{Re}\bra{f}\op{U}_{T-t}\op{A}\op{U}_t\ket{i}/\bra{f}\op{U}_T\ket{i}$.}
  \label{fig:vonneumann}
\end{figure}

To derive the effect of such an interaction, consider the coupling procedure in Figure~\ref{fig:vonneumann}.  Suppose that the impulsive coupling begins at a time $t$, and that the joint state of the detector-system degrees of freedom is initially a product state $\ket{d'}\ket{i'}$, where the initial detector and system states at time $t$
\begin{align}
  \ket{d'} &= \op{U}_{t-t_0}^{(D)}\ket{d}, & \ket{i'} &= \op{U}_t\ket{i}
\end{align}
may have propagated from previous states $\ket{i}$ and $\ket{d}$ that were prepared at the possibly different earlier times $t=0$ and $t=t_0$, respectively, following the independent Hamiltonian evolution
\begin{align}
  \op{U}^{(D)}_t &= \exp(-it\op{H}_D/\hbar), & \op{U}_t &= \exp(-it\op{H}_S/\hbar).
\end{align}
After the entangling interaction of Eq.~\eqref{eq:interaction}, the detector and system are allowed to again evolve freely for (potentially different) times $T-t$ and $T'-t$, respectively, after which they are independently measured projectively in the bases $\bra{x}$ and $\bra{f}$ to obtain a pair of detector-system results $(x,f)$. Back-propagating these measured states from $T'$ and $T$ to the time immediately following the impulsive coupling $t+\delta t\approx t$ produces the effective final states
\begin{align}
  \bra{x'} &= \bra{x}\op{U}^{(D)}_{T'-t}, & \bra{f'} &= \bra{f}\op{U}_{T-t}.
\end{align}
The joint probability distribution for the results $(x,f)$ can then be written in the compact (scattering) form
\begin{align}\label{eq:jointvn}
  p_{x,f} &= \abs{\bra{x',f'}\op{V}_{DS}\ket{d',i'}}^2.
\end{align}

Importantly, changing the durations of time $T$ and $T'$ before measuring the system and detector does not affect the general form of the joint distribution in Eq.~\eqref{eq:jointvn}; only the back-propagated states $\ket{x'}$ and $\ket{f'}$ will change from the free evolution when these durations are varied.  For a concrete example of this effect, in the optical case of \cite{Ritchie1991,Kocsis2011,Lundeen2011} the free evolution of the transverse momentum (detector) produces diffraction effects after the birefringent crystal, which alters the effective back-propagated state implied by a later transverse position measurement.

For a sufficiently small coupling strength $g$, the interaction only \emph{weakly} perturbs the initial system states, and we can expand the joint interaction $\op{V}_{DS}$ perturbatively. To good approximation, we find that the relative change of the joint probability in Eq.~\eqref{eq:jointvn} due to the weak interaction has the form
\begin{align}\label{eq:jointwv}
  \frac{p_{x,f}}{\abs{\braket{x',f'}{d',i'}}^2} &\approx 1 + 2g\,\text{Im}F_wA_w + g^2\,|F_w|^2|A_w|^2,
\end{align}
which involves only the \emph{first-order} (complex) detector and system weak values 
\begin{align}
  F_w &= \frac{\bra{x'}\op{F}\ket{d'}}{\braket{x'}{d'}}, & A_w &= \frac{\bra{f'}\op{A}\ket{i'}}{\braket{f'}{i'}}.
\end{align}
Note that continuing this expansion will produce an infinite series characterized entirely by higher-order weak values involving all powers of $\op{A}$ and $\op{F}$ \cite{Dressel2014}; however, this truncation that involves only the first-order weak values is remarkably accurate for sufficiently small coupling strengths that satisfy $g|F_w||A_w| \ll 1$ \cite{DiLorenzo2012}.

This joint distribution also determines the relative change of the marginalized distributions for the detector, $p_x = \sum_f\,p_{x,f}$, and system, $p_f = \sum_x\,p_{x,f}$, statistics alone
\begin{align}\label{eq:detectorwv}
 \frac{p_{x}}{\abs{\braket{x'}{d'}}^2} &\approx 1 + 2g\,\mean{A}\text{Im}F_w + g^2\,\mean{A^2}|F_w|^2, \\
 \frac{p_{f}}{\abs{\braket{f'}{i'}}^2} &\approx 1 + 2g\,\mean{F}\text{Im}A_w + g^2\,\mean{F^2}|A_w|^2.
\end{align}
Critically, note that the reduced \emph{detector} statistics involve the expectation value of the \emph{system} operator $\mean{A} = \bra{i'}\op{A}\ket{i'}$, as well as its second moment $\mean{A^2} = \bra{i'}\op{A}^2\ket{i'}$.  This dependence means that by examining only the statistics of the detector, we can \emph{indirectly estimate} the system observable $\op{A}$.  Similarly, the system statistics involve the expectation value of the detector operator $\mean{F} = \bra{d'}\op{F}\ket{d'}$, as well as its second moment $\mean{F^2} = \bra{d'}\op{F}^2\ket{d'}$, so can be used to indirectly estimate the detector observable $\op{F}$ in a symmetric way.

To perform such an estimation using only the detector statistics in the linear response regime (where we can neglect the terms that are second-order in $g$), an experimenter weights each of the outcomes $x$ of the detector by some scaling value $\alpha_x$, which constructs an effective detector readout observable \cite{Dressel2010,Kofman2012}
\begin{align}
  \op{R} &= \textstyle{\sum_x} \alpha_x\,\pr{x},
\end{align}
and produces the detector average (keeping all system results)
\begin{align}\label{eq:estimate}
  \textstyle{\sum_{x}}\,\alpha_x\,p_{x} &\approx \mean{R} + 2g\, \mean{A}\text{Im}\mean{RF},
\end{align}
in terms of the detector quantities $\mean{R} = \bra{d'}\op{R}\ket{d'}$ and $\mean{RF} = \bra{d'}\op{R}\op{F}\ket{d'}$.  Hence, by calibrating a known initial detector state $\ket{d'}$ and choosing the observables $\op{F}$ and $\op{R}$ strategically, one can extract the expectation value $\mean{A}$ in an unbiased way using only the detector statistics.  Note that typically the detector state $\ket{d'}$ is chosen such that the relevant observables have zero mean prior to the interaction, $\mean{F} = \mean{R} = 0$. 

Now suppose our experimental setup additionally filters the system outcomes so that only $f$ may occur.  (Alternatively, we can select only those particular events in the post-processing of data that includes more outcomes.) The statistics of the laboratory detector $x$ measurements that are properly conditioned on a particular system $f$ outcome will then have the usual form from Bayes' theorem
\begin{align}
  p_{x|f} &= \frac{p_{x,f}}{\textstyle{\sum_x} p_{x,f}}.
\end{align}
The approximate relative change of this conditional distribution is thus
\begin{align}\label{eq:wvmeas}
  \frac{p_{x|f}}{\abs{\braket{x'}{d'}}^2} &\approx \frac{1 + 2g\,\text{Im}F_wA_w + g^2\,|F_w|^2|A_w|^2}{1 + 2g\,\mean{F}\text{Im}A_w + g^2\,\mean{F^2}|A_w|^2}.
\end{align}
When the terms of order $g^2$ can be neglected, and when $\mean{F}=0$, we thus have the linear response result that can be compared to the unfiltered estimation in Eq.~\eqref{eq:estimate}
\begin{align}\label{eq:condestimate}
  \textstyle{\sum_x}\,\alpha_x\,p_{x|f} &\approx \mean{R} + 2g\, \text{Im}A_w\mean{RF}, \\
  &= \mean{R} + 2g\, [\text{Re}A_w\text{Im}\mean{RF} + \text{Im}A_w\text{Re}\mean{RF}]. \nonumber
\end{align}
Importantly, the weak value factor $\text{Re}A_w$ that scales $\text{Im}\mean{RF}$ corresponds directly to the expectation value $\mean{A}$ in Eq.~\eqref{eq:estimate}.  That is, the filtering procedure partitions the total average $\mean{A}$ into subensembles that have conditioned averages of $\text{Re}A_w$, which precisely matches what we expect from the best estimate of $\op{A}$ in Eq.~\eqref{eq:wv}.  The final term involving $\text{Im}A_w$ averages to zero in the total ensemble, and corresponds to the intrinsic symmetric backaction of the detector on the system due to the joint interaction of Eq.~\eqref{eq:interaction}; it does not correspond to the estimation of $\op{A}$ \cite{Steinberg1995,Steinberg1995b,Dressel2012d}, and can be removed from the detector signal in practice while preserving the estimation of $\mean{A}$ by choosing the detector observables such that $\text{Re}\mean{RF} = 0$.

\subsection{Observable estimation}
The preceding discussion of the von Neumann coupling is traditional in the weak value literature \cite{Aharonov1988,Dressel2014,Kofman2012} and is often sufficient for describing experimental implementations that measure weak values (e.g., \cite{Ritchie1991,Kocsis2011,Lundeen2011}).  However, this perturbative derivation makes the association between the estimations of the total observable average $\mean{A}$ and the real part of the weak value $\text{Re}A_w$ somewhat inferential, prompting skepticism about the appropriateness of the connection.  Since $\text{Re}A_w$ has such counterintuitive properties, having a more direct link between the estimation of $\op{A}$ and $\text{Re}A_w$ as a conditioned average value more generally is desirable.  Thankfully, we can indeed demonstrate that this is the proper association by rephrasing the conditional estimation procedure using the formalism of generalized measurements \cite{Dressel2010,Dressel2012b,Dressel2013b}.

To do this, we rewrite the joint probability of Eq.~\eqref{eq:jointvn} entirely in the system space
\begin{align}\label{eq:joint}
  p_{x,f} = \abs{\bra{f'}\op{M}_x\ket{i'}}^2,
\end{align}
by defining a \emph{measurement (Kraus) operator} that encodes the entire coupling and measurement procedure into a single system operator \cite{Nielsen2000,Wiseman2009}
\begin{align}
  \op{M}_x &= \bra{x'}\op{V}_{DS}\ket{d'}.
\end{align}
Notice that this measurement operator has the form of a partial matrix element, but only contracts out the detector part of the joint unitary $\op{V}_{DS}$ to leave a purely system operator, which directly corresponds to how the complete measurement \emph{procedure} affects the system. Intuitively, for strong measurements of $\op{A}$, the measurement operators $\op{M}_x \to \pr{a}$ will become projectors onto the eigenstates of $\op{A}$, while for \emph{weak} measurements (in the sense of \cite{Aharonov1988}) the measurement operators $\op{M}_x \approx \op{1}$ will approximate the identity operator for all $x$, which leaves the initial state nearly unperturbed.

It follows that the marginalized distribution of only the detector results can be written
\begin{align}
  p_x &= \textstyle{\sum_f}\, p_{x,f} = \bra{i'}\op{M}_x^\dagger\op{M}_x\ket{i'},
\end{align}
which has the form of the \emph{system} expectation of a probability operator
\begin{align}
  \op{P}_x &= \op{M}^\dagger_x\op{M}_x
\end{align}
for the \emph{detector} result $x$.  These probability operators are positive, and form a resolution of the identity in the system space
\begin{align}\label{eq:pom}
  \textstyle{\sum_x} \op{P}_x = \bra{d'}\op{V}_{DS}^\dagger\left[\textstyle{\sum_x} \pr{x'}\right]\op{V}_{DS}\ket{d'} = \op{1},
\end{align}
making them a \emph{probability operator-valued measure} (POM, or POVM).  Such a POM is the operator version of a properly normalized probability distribution.  Indeed, when $\op{P}_x$ commutes with $\op{A}$, its diagonal elements will be precisely the classical conditional probabilities $p_{x|a}$ describing the likelihoods of each detector result $x$ given a definite preparation of $a$ \cite{Dressel2012b}.  For each $a$, these probabilities then independently satisfy $\sum_x p_{x|a} = 1$ according to Eq.~\eqref{eq:pom}.

Generally speaking, any purity-preserving generalized measurement can be expressed as such a measurement operator $\op{M}_x$ and associated POM $\op{P}_x = \op{M}_x^\dagger\op{M}_x$ \cite{Nielsen2000,Wiseman2009}, with the von Neumann interaction being a special case for implementing such a measurement with a concrete Hamiltonian. For example, the polarization-dependent reflection measurement in \cite{Dressel2011} did not use a von Neumann Hamiltonian description of the interaction of the entangled photon pair with the glass microscope coverslip in the experimental analysis, but rather characterized the relevant POM operators $\op{P}_x$ directly with a series of separate calibration measurements by \emph{measuring} the conditional probabilities $p_{x|a}$ for known preparations of the eigenstates $\ket{a}$. Up to additional phases that were also calibrated in separate measurements, the diagonal elements of $\op{M}_x$ then had the form $\sqrt{p_{x|a}}$. The benefit of this direct approach is that the effect of the actual measurement procedure may be experimentally measured, without requiring a more detailed model of the extended detector space.

Now suppose that we can use the measured probabilities $p_x$ to estimate the expectation value of $\op{A}$ in an unbiased way. To do this, we must weight the outcomes $x$ of the detector with appropriately scaled values $\alpha_x$ (e.g., by rescaling a low-visibility signal in the usual way) \cite{Dressel2010} 
\begin{align}\label{eq:expect}
  \textstyle{\sum_x} \alpha_x\,p_x &= \bra{i'}[\textstyle{\sum_x} \alpha_x\, \op{P}_x]\ket{i'}.
\end{align}
Evidently, to produce the expectation value $\mean{A}=\bra{i'}\op{A}\ket{i'}$ for any initial state $\ket{i'}$, we must be able to choose appropriate values $\alpha_x$ that will calibrate the detector to probe the generalized spectral expansion \cite{Dressel2010,Dressel2012b}  
\begin{align}\label{eq:identity}
  \op{A} = \textstyle{\sum_x} \alpha_x \op{P}_x.
\end{align}
If it can be arranged, this operator identity will guarantee that the values $\alpha_x$ and measurement operators $\op{M}_x$ (and thus the probability operators $\op{P}_x$) will directly estimate $\op{A}$ in an unbiased way.  As a concrete example, in \cite{Aharonov1988} the detector was chosen such that $\op{F} = \op{p}$ was the momentum operator, $\op{R} = \op{x}$ was the position operator, and $\braket{x}{d'} = (2\pi\sigma^2)^{-1/4}\exp(-x^2/4\sigma^2)$ was a zero-mean Gaussian distribution, such that $\mean{R}=\mean{F}=0$, $F_w = ix/2\sigma^2$, and thus $\mean{\op{R}\op{F}} = i/2$ in Eq.~\eqref{eq:estimate}; setting the simple scaled values of $\alpha_x = x/g$ then directly yields an unbiased estimation of $\mean{A}$ for all $g$. Note, however, that this common choice for Gaussian measurements is generally not a unique choice for producing an unbiased estimation, since the dimension of the detector often exceeds that of the system \cite{Dressel2010,Dressel2012b}.

Once we fix the weights $\alpha_x$ to achieve the estimation identity of Eq.~\eqref{eq:identity}, we also fix the partial average for each postselection $f$ according to Eq.~\eqref{eq:joint}
\begin{align}\label{eq:partial}
  \textstyle{\sum_x} \alpha_x\,p_{x,f} &= \bra{i'}\op{O}_t\ket{i'}, 
\end{align}
which we write compactly as an expectation value for an effective system operator at time $t$
\begin{align}
  \op{O}_t &\equiv \textstyle{\sum_x}\alpha_x \op{M}^\dagger_x\op{\Pi}_f\op{M}_x, \\
  &= \frac{1}{2}(\op{A}\op{\Pi}_f + \op{\Pi}_f\op{A}) + \textstyle{\sum_x} \alpha_x\,\mathcal{L}[\op{M}^\dagger_x]\op{\Pi}_f, \nonumber
\end{align}
that we write in turn as a symmetric (Jordan) product \cite{Jordan1934} between the operator $\op{A}$ and the postselection projection operator $\op{\Pi}_f \equiv \pr{f'}$, modified by a sum of weighted Lindblad (dissipation) operations \cite{Dressel2013b}
\begin{align}\label{eq:lindblad}
  \mathcal{L}[\op{M}^\dagger_x](\cdot) &\equiv \frac{1}{2}\left([\op{M}^\dagger_x,\cdot]\op{M}_x + \op{M}^\dagger_x[\cdot,\op{M}_x]\right),
\end{align}
familiar from open-system dynamics \cite{Wiseman2009,Breuer2007}. These Lindblad terms quantify the perturbation introduced by the measurement. Note that for \textit{weak} measurements (i.e., $\op{M}_x \approx \op{1}$) the commutators in the Lindblad terms approximately vanish to leave only the symmetric product with $\op{A}$, signifying that both the initial system state and postselection are essentially unaffected by the measurement results $x$ that are being used to estimate $\op{A}$ \cite{Dressel2012b,Dressel2013b}.

Expanding the partial average in Eq.~\eqref{eq:partial} produces
\begin{align}\label{eq:partial2}
  \textstyle{\sum_x} \alpha_x\,p_{x,f} &= \text{Re}\,\bra{f'}\op{A}\ket{i'}\braket{i'}{f'} + \mathcal{E}[\alpha],
\end{align}
where we have introduced the Lindblad error terms
\begin{align}
  \mathcal{E}[\alpha] = \textstyle{\sum_x} \alpha_x \bra{i'}\,(\mathcal{L}[\op{M}_x]\op{\Pi}_f)\,\ket{i'}
\end{align}
that are produced entirely by the perturbation from the measurement.  Conditioning this partial average on obtaining a particular $f$ then yields
\begin{align}\label{eq:condav}
  \frac{\sum_x \alpha_x\,p_{x,f}}{\sum_x\,p_{x,f}} &= \frac{\text{Re}\,\bra{f'}\op{A}\ket{i'}\braket{i'}{f'} + \mathcal{E}[\alpha]}{\braket{f'}{i'}\braket{i'}{f'} + \mathcal{E}[1]}.
\end{align}
When the error terms $\mathcal{E}$ are small enough to be neglected \cite{Dressel2012,Dressel2012b} (meaning that the initial system state is negligibly perturbed), the \emph{real part} of the weak value in Eq.~\eqref{eq:wv} is unambiguously recovered as the \textit{measured} conditioned estimate for $\op{A}$, verifying our derivation of this real part as a best estimate. As expected, the imaginary part is unrelated to the estimation of $\op{A}$, so does not contribute to Eq.~\eqref{eq:condav}, which justifies our interpretation of the terms in the von Neumann linear response of Eq.~\eqref{eq:condestimate}. Deriving Eq.~\eqref{eq:wvgen} as a measured estimation is a similar exercise \cite{Dressel2012e}.

Importantly, nothing about the derivation of Eq.~\eqref{eq:condav} changes when the time $t$, the initial system state $\ket{i}$, the system postselection $\bra{f}$, or even the system Hamiltonian $\op{H}$ are varied, as long one keeps fixed the measurement procedure set by the choice of calibration weights $\alpha_x$ and corresponding $\op{M}_x$ (e.g., the detector states $\ket{d'}$ and $\bra{x'}$, Hamiltonian $\op{H}^{(D)}$, and coupling interaction $\op{V}_{DS}$). This robustness of the derivation implies that the same weak measurement procedure can approximate the \emph{entire functional dependence} of the weak value in Eq.~\eqref{eq:wv}, in contrast to the single arbitrary value produced by the coin disturbance scheme in Ref.~\cite{Ferrie2014b}.  Moreover, the weak value in Eq.~\eqref{eq:wv} no longer depends upon the specific measurement procedure, just like the expectation value in Eq.~\eqref{eq:expect}, so this approximation will work for \emph{any} unbiased weak measurement procedure. The only requirement for consistently recovering the weak value Eq.~\eqref{eq:wv} as the limiting value of the conditioned average in Eq.~\eqref{eq:condav} is for the Lindblad perturbation terms in Eq.~\eqref{eq:lindblad} to be small enough to neglect \cite{Dressel2012e,Dressel2013b}, meaning that the quantum state is approximately unperturbed. (For a physical example where the state disturbance from the coupling may not always be neglected, see \cite{Dressel2012c}.).

\subsection{Dynamical weak values}
\begin{figure}[t]
  \includegraphics[width=0.9\columnwidth]{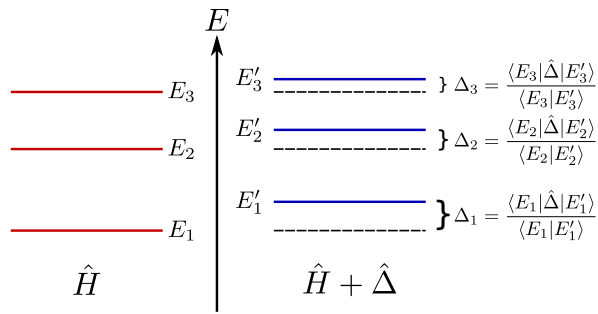}
  \caption{Energy level perturbations.  The eigenvalues $E$ of a Hamiltonian $\op{H}$ are shifted to new eigenvalues $E'$ when a Hamiltonian perturbation $\op{\Delta}$ is added.  These shifts in energy levels are exactly (real) weak values $\bra{E}\op{\Delta}\ket{E'}/\braket{E}{E'}$, and thus may lie outside the spectrum of the perturbation $\op{\Delta}$.}
  \label{fig:perturb}
\end{figure}

Thus far we have carefully discussed the most well-known role of weak values in experiment, namely as complex parameters that characterize a von Neumann interaction, and as conditioned estimations of observable averages.  However, these are not the only places that weak values naturally appear.  Here we consider two more common cases that are usually overlooked: weak values as eigenvalue perturbations, and weak values as classical dynamical variables in reduced system dynamics.

The first case is mathematically trivial to show, but has nontrivial implications.  Consider the case of a Hamiltonian $\op{H}$ with energy eigenstates $\ket{E}$ and eigenvalues $\op{H}\ket{E} = E\ket{E}$.  Suppose this Hamiltonian becomes perturbed by a new contribution $\op{\Delta}$, producing new eigenstates $(\op{H}+\op{\Delta})\ket{E'} = E'\ket{E'}$.  This latter eigenvalue equation can then be contracted with an unperturbed eigenstate $\bra{E}$ and rearranged to find the following relation between the eigenvalues:
\begin{align}
  E' &= E + \frac{\bra{E}\op{\Delta}\ket{E'}}{\braket{E}{E'}},
\end{align}
which is illustrated in Figure~\ref{fig:perturb}. That is, the (purely real) weak value of the perturbation $\op{\Delta}$ determines the shift in energy for the eigenstates of the Hamiltonian, and thus may lie outside the spectrum of $\op{\Delta}$.  One can understand this weak value as the best estimate of the average energy perturbation required to move from the old eigenstate $\ket{E}$ to the new eigenstate $\ket{E'}$.  No measurement is being performed here, so this shift constitutes a dynamical effect where the weak value indeed represents the physical energy shift.  This shift can be verified by measuring the eigenenergies before and after such a perturbation is added to the system.

\begin{figure}[t]
  \includegraphics[width=\columnwidth]{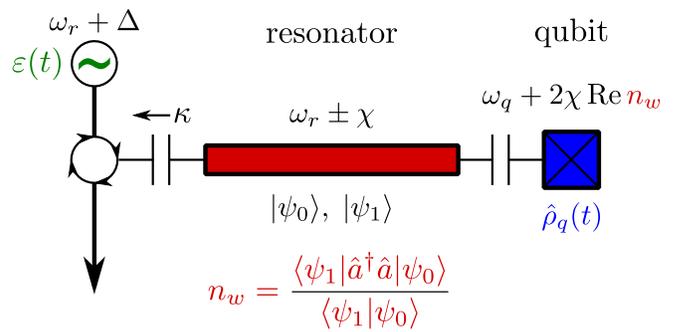}
  \caption{Superconducting qubit capacitively coupled to a one-sided stripline resonator with energy-decay rate $\kappa$, which is pumped through a circulator with a coherent microwave source $\varepsilon(t)$ at a frequency detuned by $\Delta$ from the bare resonator frequency $\omega_r$. The resonator frequency is shifted by $\pm \chi$ depending on the qubit state due to the dispersive coupling. As such, each definite qubit state $\ket{0}$ or $\ket{1}$ approximately correlates to a distinct resonator state $\ket{\psi_0}$ or $\ket{\psi_1}$. The reduced qubit state $\op{\rho}_q(t)$ then displays coherence oscillations at a frequency $\omega_q + 2\chi\, \text{Re}\,n_w(t)$ due to the ac Stark shift, which depends on the real part of the weak value $n_w(t) = \bra{\psi_1}\op{a}^\dagger\op{a}\ket{\psi_0}/\braket{\psi_1}{\psi_0}$ of the resonator population.  The qubit coherence similarly displays decay at an average rate $\Gamma = 2\chi\,\text{Im}\,n_w(t)$ that depends on the imaginary part of $n_w(t)$, indicating measurement-dephasing from ensemble-averaging the fluctuations of the resonator population. Importantly, the complex weak value $n_w(t)$ is \emph{physically} the relevant mean-field classical dynamical variable for the resonator population that affects the reduced qubit state at every point in time $t$, with the real part corresponding to the best classical estimation of the ensemble-averaged resonator population probed by coherent qubit superpositions of $\ket{0}$ and $\ket{1}$.}
  \label{fig:cqed}
\end{figure}

The second case of a dynamical weak value is best illustrated by an explicit example, shown in Figure~\ref{fig:cqed}.  Suppose we couple a qubit dispersively to a single-mode resonator (e.g., a superconducting qubit setup like the one used in \cite{Groen2013}).  The simplest Hamiltonian for how such a joint system naturally evolves is
\begin{align}\label{eq:qubitham}
  \op{H} &= \frac{\hbar\,\omega_q}{2}\,\op{\sigma}_z + \hbar\,\omega_r\, \op{a}^\dagger\op{a} + \hbar\,\chi\,\op{\sigma}_z\,\op{a}^\dagger\op{a},
\end{align}
where $\op{\sigma}_z = \pr{1} - \pr{0}$ is the Pauli Z-operator between the qubit energy levels, $\op{a}$ is the lowering operator of the resonator mode satisfying $[\op{a},\op{a}^\dagger]=1$, $\omega_q$ and $\omega_r$ are the oscillation frequencies of the qubit and resonator, and $\pm \chi$ is the dispersive frequency shift of the resonator that depends on the qubit state.  Note that the interaction term between the qubit and the resonator has the general von Neumann form of Eq.~\eqref{eq:vnham}, but it is no longer impulsive.

Assuming a pure state for the joint qubit-resonator system, we can make the following ansatz for the form of the joint state
\begin{align}\label{eq:ansatz}
  \ket{\Psi} &= c_0(t)\,\ket{0}\ket{\psi_0(t)} + c_1(t)\,\ket{1}\ket{\psi_1(t)},
\end{align}
where $c_{0,1}(t)$ are complex amplitudes, and $\ket{\psi_{0,1}(t)}$ are normalized resonator states that are correlated to each definite qubit state.  It follows that the reduced qubit density matrix $\op{\rho}_q = \text{Tr}_r\pr{\Psi}$ after tracing out the resonator has the following diagonal populations
\begin{align}\label{eq:qubitpops}
  p_0(t) &= |c_0(t)|^2, & p_1(t) &= |c_1(t)|^2,
\end{align}
as well as an off-diagonal coherence
\begin{align}\label{eq:qubitcoh}
  \rho_{01}(t) &= c_1^*(t)c_0(t)\braket{\psi_1(t)}{\psi_0(t)},
\end{align}
that depends explicitly on the overlap between the two distinct and dynamically evolving resonator states that are correlated to definite qubit populations.

The simple Hamiltonian considered in Eq.~\eqref{eq:qubitham} is already diagonal in the energy basis of the qubit, so the populations in Eq.~\eqref{eq:qubitpops} do not change in time.  However, the coherence in Eq.~\eqref{eq:qubitcoh} will display phase oscillations due to both the natural qubit energy splitting and the added influence of the dispersive resonator coupling.  After extracting the components of the joint Schr\"odinger equation $i\hbar\partial_t \ket{\Psi} = \op{H}\ket{\Psi}$ by differentiating Eq.~\eqref{eq:ansatz}, and a bit of algebra, it is straightforward to derive the evolution equation for the coherence directly from differentiating Eq.~\eqref{eq:qubitcoh}
\begin{align}\label{eq:coherence}
  \partial_t \rho_{01}(t) &= i[\omega_q + 2\chi\, n_w(t)]\,\rho_{01}(t).
\end{align}
Notably, the (complex) weak value of the resonator population naturally appears
\begin{align}\label{eq:resonatorwv}
  n_w(t) &= \frac{\bra{\psi_1(t)}\op{a}^\dagger\op{a}\ket{\psi_0(t)}}{\braket{\psi_1(t)}{\psi_0(t)}} = \bar{n}(t) + i\bar{n}_\gamma(t).
\end{align}
This weak value can be understood as a \emph{classical dynamical variable} that completely determines the ensemble-averaged \emph{dynamical} influence of the resonator on the reduced qubit state at each point in time $t$, even in the absence of any explicit preselection or postselection measurements.  

The real part $\bar{n}(t)$ of this weak value is the best estimate of the population of the resonator mode if the qubit transitions between its ground and excited states at time $t$.  According to Eq.~\eqref{eq:coherence}, this weak value produces a shift $2\chi\bar{n}(t)$ of the natural qubit frequency, commonly known as the \emph{ac Stark shift} \cite{Bonch-Bruevich1969,Brune2004,Schuster2005}.  Notably, this shift does not involve either of the average mode populations $n_0 = \bra{\psi_0}\op{a}^\dagger\op{a}\ket{\psi_0}$ or $n_1 = \bra{\psi_1}\op{a}^\dagger\op{a}\ket{\psi_1}$ that one might naively expect, since the qubit coherence does not pertain to a definite population.  Any imaginary part $\bar{n}_\gamma(t)$ of the weak value does not contribute to the ac Stark shift in Eq.~\eqref{eq:coherence}, but instead produces a decay of the qubit coherence that indicates the resonator coupling is \emph{dephasing} the qubit at a rate $2\chi\bar{n}_\gamma(t)$.

For an explicit example of this reduced evolution, to a good approximation \cite{Gambetta2006,Gambetta2008,Boissonneault2009} a one-sided resonator with energy-decay rate $\kappa$ (as shown in Figure~\ref{fig:cqed}) that is pumped with a coherent state $\varepsilon$ detuned by $\Delta$ from the bare resonator frequency will reach a steady state that approximates the pure-state ansatz of Eq.~\eqref{eq:ansatz}, with $\ket{\psi_0}$ and $\ket{\psi_1}$ approximating coherent states with complex classical amplitudes
\begin{align}
  \psi_0 &\equiv \bra{\psi_0}\op{a}\ket{\psi_0} = \frac{2\varepsilon}{\kappa}\frac{1}{1 + i2(\Delta - \chi)/\kappa}, \\
  \psi_1 &\equiv \bra{\psi_1}\op{a}\ket{\psi_1} = \frac{2\varepsilon}{\kappa}\frac{1}{1 + i2(\Delta + \chi)/\kappa}.
\end{align}
The weak value of the resonator population in Eq.~\eqref{eq:resonatorwv} correspondingly has the simple steady-state form (assuming a wide-bandwidth resonator for brevity, with $\chi,\Delta \ll \kappa$)
\begin{align}
  n_w = \psi_1^*\psi_0 \approx \frac{4\varepsilon^2}{\kappa^2}\left[1 + i\frac{4\chi}{\kappa}\right],
\end{align}
which produces the ac Stark shift $2\chi\bar{n}$ of the qubit frequency with $\bar{n} = 4\varepsilon^2/\kappa^2$, as well as the dephasing rate $\Gamma = 2\chi\bar{n}_\gamma = 8\chi^2\bar{n}/\kappa$. These expressions for the ac Stark shift and ensemble-average dephasing rate agree with the known results for dispersive qubit measurements in circuit quantum electrodynamics (cQED) \cite{Gambetta2006,Gambetta2008,Boissonneault2009,Korotkov2011}.

We emphasize that the complex weak value of the resonator population in Eq.~\eqref{eq:resonatorwv} is \emph{physically} the relevant classical dynamical variable that controls the behavior of the ensemble-averaged reduced qubit evolution in Eq.~\eqref{eq:coherence}.  Indeed, the ac Stark shift of the qubit frequency is the primary method used in cQED for extracting the average population $\bar{n}$ in the resonator at steady state (e.g., \cite{Jeffrey2014}), but we see here that in actuality such a dynamical method probes the \emph{weak value} of that population, and not the population associated with any particular qubit state, or even the total average population $\bar{n}_{\text{tot}} = \bra{\Psi}\op{a}^\dagger\op{a}\ket{\Psi}$. The weak value arises from the interference between the two fields $\ket{\psi_0}$ and $\ket{\psi_1}$ in the resonator that are correlated with the two definite qubit states, on average. 

At each time $t$ the real part of $n_w$ indicates the best estimate of the average resonator excitation number seen by a qubit that is not in a definite energy eigenstate, while its imaginary part indicates the best estimate of the backaction on the qubit dynamics caused by the fluctuations of the resonator population around that average value.  These latter fluctuations result in ensemble-average dephasing of the reduced qubit state, reinforcing the observation made for weak measurements that a weak value can only describes the average (i.e., classical mean field) state of affairs for an ensemble of realizations \cite{Dressel2014b}. Indeed, when the time-dependent leakage from the resonator is accounted for, the qubit will not dephase in this manner, but will instead follow a pure \emph{quantum trajectory} \cite{Gambetta2008,Korotkov2011,Murch2013,Weber2014} that depends on the leakage record. As such, we must necessarily interpret the weak value $n_w$ here as implicitly averaging over many such realizations in practice to produce a classical dynamical variable associated with the classical mean field in the resonator.

\section{Disturbance, quasiprobabilities, and Hamilton-Jacobi}\label{sec:quasiprob}
Given the consistent role of the real part of a weak value as a best conditioned observable estimate, we can now observe an intriguing logical tension inherent to the weak value. On one hand, any classical conditioned average must include disturbance to obtain anomalous values \cite{Williams2008,Dressel2011,Dressel2012b,Tollaksen2007,Ipsen2014}: the larger the disturbance, the more strange the average can become. On the other hand, the strangeness of the conditioned average in Eq.~\eqref{eq:condav} is greatest when the quantum state is \emph{least} disturbed by an intermediate measurement \cite{Dressel2012e}, and even persists when there is no added disturbance to the natural dynamical evolution in the example of Eq.~\eqref{eq:coherence}. 

These two statements imply that any classical (hidden-variable) explanation of a strange weak value as a disturbance effect must satisfy one of two properties: either (a) the quantum state must be a \textit{subjective} (epistemic) quantity that is completely insensitive to whatever physical (ontic) disturbance is occurring, or (b) the relevant disturbance occurs entirely during the \textit{postselection}, and not the intermediate measurement \cite{Dressel2012b}. 

Classical fields that produce strange weak values in weak measurement experiments satisfy this second property, where the disturbance at the postselection filter causes interference between previously independent field components \cite{Ritchie1991,Bliokh2013a,Bliokh2014,Dressel2014b}. However, quantum systems can display similar interference without wavelike intensities \cite{Pryde2005,Goggin2011,Groen2013}, and permit additional entanglement effects \cite{Dressel2011} such as the dynamical evolution involving weak values emphasized in the previous section. Especially for such a dynamical role of the weak value, it does not seem possible to ascribe strange weak values to classical disturbance mechanisms without also demanding that the quantum state is itself a fundamentally subjective collection of some more physical microstates; such a demand, while not impossible, must contend with the Pusey-Barrett-Rudolph theorem \cite{Pusey2012} and the other no-go theorems (reviewed in \cite{Leifer2014}) for states of such epistemic character. 

\subsection{Terletsky-Margenau-Hill and Kirkwood-Dirac}
To quantify this logical tension, we can express weak values in a more established and familiar way by rewriting Eq.~\eqref{eq:wv} using the spectral expansion $\op{A} = \sum_a \,a\,\pr{a}$ to find $A_w(t) = \sum_a\,a\,\tilde{p}_{a|i,f}$, where
\begin{align}\label{eq:quasicond}
  \tilde{p}_{a|i,f} &= \frac{\text{Re}\,\braket{f'}{a}\braket{a}{i'}\braket{i'}{f'}}{|\braket{f'}{i'}|^2}.
\end{align}
This is a conditional \textit{quasiprobability} distribution that weights the eigenvalues of $\op{A}$ in $A_w$, and satisfies the normalization $\sum_a \tilde{p}_{a|i,f} = 1$.  As a result, if a strange weak value $|A_w| > ||\op{A}||$ is estimated, then at least one quasiprobability must be negative: $\tilde{p}_{a|i,f} < 0$.  

Since the conditioning denominator of Eq.~\eqref{eq:quasicond} is positive-definite, we infer that the joint quasiprobability
\begin{align}\label{eq:kirkwood}
  \tilde{p}_{a,f|i} &= \text{Re}\,\braket{f'}{a}\braket{a}{i'}\braket{i'}{f'}
\end{align}
in the numerator of Eq.~\eqref{eq:quasicond} [also appearing directly in the experimental partial average of Eq.~\eqref{eq:partial2}] must be negative. This joint quasiprobability distribution is precisely the \emph{Terletsky-Margenau-Hill distribution} \cite{Terletsky1937,Margenau1961,Johansen2004,Johansen2004c} that has been used since the late 1930s.

Interestingly, the Terletsky-Margenau-Hill distribution is the real part of the complex quasiprobability distribution introduced even earlier by Kirkwood \cite{Kirkwood1933,Dirac1945,Chaturvedi2006,Lundeen2011,Lundeen2012,Hofmann2012b,Lundeen2014} as an alternative to the Wigner distribution \cite{Wigner1932}. This Kirkwood distribution is also known as the \emph{standard-ordering distribution} for quantum phase space \cite{Mehta1964}. In fact, Dirac later considered this distribution specifically to discuss the classical-to-quantum transition \cite{Dirac1945}, observing that the negativity arises directly from the usual operator noncommutativity of quantum mechanics. Notably, the fully complex weak value that appears in reduced state dynamics is nothing more than a conditioned version of this complex Kirkwood-Dirac quasiprobability distribution. 

An important feature of the Kirkwood-Dirac distribution that has recently come to light \cite{Lundeen2011,Lundeen2012} is that any quantum state can be written in an operator basis such that this distribution forms its components. That is, if we define a suitable operator basis 
\begin{align}
  \Gamma_{a,f} = \frac{\ket{a}\bra{f}}{\braket{f}{a}}
\end{align}
then we can write any quantum state using the expansion 
\begin{align}
  \op{\rho} = \sum_{a,f} \,\braket{f}{a}\bra{a}\op{\rho}\ket{f}\,\Gamma_{a,f}.
\end{align}
As such, the quasiprobabilities $\rho(f,a) = \braket{f}{a}\bra{a}\op{\rho}\ket{f}$ of the complex Kirkwood-Dirac distribution are a complete quantum state representation for an arbitrary density matrix that is analogous to a complex wavefunction $\psi(x) = \braket{x}{\psi}$ for a pure state. Unlike the usual wavefunction, however, the Kirkwood-Dirac distribution is directly compatible with Bayes' theorem [as used in Eq.~\eqref{eq:quasicond}], and thus behaves like a true probability distribution. This notable feature has enabled alternative methods of quantum state tomography by directly measuring complex weak values using von Neumann interactions \cite{Lundeen2011,Lundeen2012,Salvail2013,Malik2014,Howland2014,Mirhosseini2014,Hofmann2014}, and also permits fully Bayesian quasiprobabilistic reformulations of coherent quantum dynamics \cite{Hofmann2012b,Lundeen2014}.

\subsection{Wigner distribution and negativity}
Importantly, the negativity in such a quasiprobability representation of a quantum state is closely associated to traditional measures of nonclassicality \cite{Spekkens2008,Ferrie2011}. The usual examples of this criterion for nonclassicality are the negativity of the Wigner distribution \cite{Wigner1932}, or the Glauber-Sudarshan $P$-distribution \cite{Glauber1963,Sudarshan1963}; however, the Terletsky-Margenau-Hill distribution in Eq.~\eqref{eq:kirkwood} has the same feature \cite{Johansen2004c}.

To emphasize this point, we can in fact relate weak values directly to the Wigner distribution when we are considering infinite-dimensional systems. To see this, consider the Wigner distribution for an initially pure state $\ket{i}$ \cite{Wigner1932}
\begin{align}
  W_i(x,p) &= \int_{-\infty}^\infty\!\! \braket{x-\frac{y}{2}}{i}\braket{i}{x+\frac{y}{2}}e^{ipy/\hbar}\frac{dy}{2\pi\hbar},
\end{align}
where $x$ and $p$ represent the usual classical position and momentum variables. Now suppose we compute the partial average of the momentum $p$ for a fixed $x$ using this joint quasiprobability distribution
\begin{align}\label{eq:wignerpartial}
  &\int_{-\infty}^\infty\!\! p\,W_i(x,p)dp = {} \\
  &\quad = \iint_{-\infty}^\infty\!\! \braket{x-\frac{y}{2}}{i}\braket{i}{x+\frac{y}{2}}(-i\hbar\partial_y e^{ipy/\hbar})\frac{dydp}{2\pi\hbar}, \nonumber \\
  &\quad = \int_{-\infty}^\infty\!\! i\hbar\partial_y \left[\braket{x-\frac{y}{2}}{i}\braket{i}{x+\frac{y}{2}}\right] \int_{-\infty}^\infty\!\! e^{ipy/\hbar}\frac{dp}{2\pi\hbar}dy, \nonumber \\
  &\quad = \int_{-\infty}^\infty\!\! \frac{(-i\hbar\partial_x\braket{x-\frac{y}{2}}{i})\braket{i}{x+\frac{y}{2}} + \text{c.c.}}{2}\, \delta(y)\,dy, \nonumber \\
  &\quad = \text{Re}(-i\hbar\partial_x\braket{x}{i})\braket{i}{x}, \nonumber \\
  &\quad = \text{Re}\bra{x}\op{p}\ket{i}\braket{i}{x}, \nonumber \\
  &\quad = \int_{-\infty}^\infty\!\! p\, \text{Re}\braket{x}{p}\braket{p}{i}\braket{i}{x}\,dp. \nonumber
\end{align}
In other words, after performing integration-by-parts the partial average of the Wigner function yields precisely the same result as averaging $p$ over the Terletsky-Margenau-Hill distribution of Eq.~\eqref{eq:kirkwood}.  It follows that conditioning this partial average directly produces a momentum weak value as the proper local average of momentum at the position $x$
\begin{align}\label{eq:wignerwv}
  \frac{\int_{-\infty}^\infty p\,W_i(x,p)dp}{\int_{-\infty}^\infty W_i(x,p)dp} = \text{Re}\frac{\bra{x}\op{p}\ket{i}}{\braket{x}{i}}.
\end{align}
As with Eq.~\eqref{eq:quasicond}, the denominator is a positive-definite probability, so only the partial average of Eq.~\eqref{eq:wignerpartial} in the numerator of Eq.~\eqref{eq:wignerwv} can produce nonclassical behavior.

The inescapable conclusion is that weak values are intimately related to quasiprobability distributions for the quantum formalism in a fundamental and unavoidable way \cite{Aharonov2005b}. Moreover, these conditioned averages consistently confirm the best estimate that we expected from the derivation of Eq.~\eqref{eq:wv}. Similar relationships can be established for the other quantum phase space Moyal distributions \cite{Moyal1949}, of which the Wigner function and the Glauber-Sudarshan $P$-distribution \cite{Glauber1963,Sudarshan1963} are special-cases. (See \cite{Bednorz2013} for an example that emphasizes this point using non-symmetrized correlation functions.)  

We therefore have the following observation: if strange conditioned averages approximate the functional dependence of the weak value in Eq.~\eqref{eq:wv}, then classical hidden variable models will be unable to satisfactorily explain that dependence. If one could, then it would also be able to reproduce other nonclassical statistical features of the quantum theory that arise from the negativity of these quasiprobability distributions (see also \cite{Tollaksen2007,Pusey2014}). Such negativity, in turn, arises from intrinsic quantum interference that is not present in classical systems of particles.

\subsection{Hamilton-Jacobi Formalism}\label{sec:hamilton}
As a last, albeit poignant, illustration of the pervasive theoretical role of weak values in the quantum mechanical formalism, let us re-examine the Schr\"odinger equation for a nonrelativistic particle
\begin{align}\label{eq:schrodinger}
  i\hbar \partial_t \ket{\psi} &= \left[\frac{\op{p}^2}{2m} + V(\op{x})\right]\ket{\psi}.
\end{align}
We can expand this equation into a wavefunction form that is local in the coordinate $x$ by contracting it with $\bra{x}$ and rearranging
\begin{align}
  \partial_t (i\hbar \ln\braket{x}{\psi}) &= \frac{1}{2m}\frac{\bra{x}\op{p}^2\ket{\psi}}{\braket{x}{\psi}} + V(x).
\end{align}
This local form automatically involves the weak values of the kinetic and potential energies.

Noting the familiar structure of this equation, we can then define a complex version of \emph{Hamilton's principle function} (i.e., the classical action) as
\begin{align}
  S(x,t) &\equiv -i\hbar \ln\braket{x}{\psi},
\end{align}
with real and imaginary parts
\begin{align}
  \text{Re}S(x,t) &= \hbar \Phi(x,t), \\
  \text{Im}S(x,t) &= -\frac{\hbar}{2}\ln \rho(x,t),
\end{align}
that naturally isolate the phase (i.e., \emph{eikonal}) $\Phi$ and probability density $\rho$ components of the wavefunction $\braket{x}{\psi} = \sqrt{\rho(x,t)}\,\exp[i\Phi(x,t)]$.

From this complex action, we can define the classically \emph{conjugate momentum field} $p(x,t)$ that corresponds to the local coordinate $x$ in the usual way
\begin{align}\label{eq:momentum}
  p(x,t) &= \partial_x S(x,t) = -i\hbar\partial_x \ln\braket{x}{\psi} = \frac{\bra{x}\op{p}\ket{\psi}}{\braket{x}{\psi}},
\end{align}
which naturally produces a complex momentum weak value as the appropriate classical dynamical variable for the local momentum field.  The real part of this weak value is the \emph{Bohmian momentum} \cite{Bohm1952a,Bohm1952b,Wiseman2007,Kocsis2011}
\begin{align}\label{eq:bohmmom}
  \text{Re}\,p(x,t) &= \hbar \partial_x \Phi(x,t),
\end{align}
while the imaginary part is the \emph{osmotic momentum} \cite{Nelson1966,Bohm1989}
\begin{align}\label{eq:osmom}
  \text{Im}\,p(x,t) &= -\frac{\hbar}{2}\partial_x \ln \rho(x,t),
\end{align}
that indicates the logarithmic change of the probability distribution $\rho(x,t) = |\braket{x}{\psi}|^2$ as $x$ is varied \cite{Dressel2012d}.  These local momentum fields jointly act as the average fluid-like momentum that corresponds to a fluctuating classical mean field for the quantum particle \cite{Madelung1926,Madelung1927}.

Using this complex action, we can thus identically rewrite Schr\"odinger's Eq.~\eqref{eq:schrodinger} as a straightforward \emph{Hamilton-Jacobi} equation for the classical mean-field $x$ and $p(x,t)$ dynamical variables \cite{Hiley2012}
\begin{align}\label{eq:hamjac}
  0 &= \partial_t S(x,t) + H[x,p(x,t),t],
\end{align}
where the classical Hamiltonian function has the form
\begin{align}
  H[x,p(x,t),t] &= \frac{1}{2m}\frac{\bra{x}\op{p}^2\ket{\psi}}{\braket{x}{\psi}} + V(x),
\end{align}
involving local position weak values of the kinetic and potential energies.  

Splitting the complex Hamilton-Jacobi equation of Eq.~\eqref{eq:hamjac} into its independent real and imaginary parts produces
\begin{align}
  \label{eq:hamjacreal}
  0 &= \hbar \partial_t \Phi(x,t) + \frac{[\text{Re}\,p(x,t)]^2}{2m} + V(x) + Q(x), \\
  \label{eq:continuity}
  0 &= \partial_t \rho(x,t) + \partial_x \left[\rho(x,t)\frac{p(x,t)}{m}\right].
\end{align}
Note that Eq.~\eqref{eq:continuity} is nothing more than the usual continuity equation for the probability density $\rho(x,t)$ in terms of the local Bohmian velocity field $v(x,t) = p(x,t)/m$, while Eq.~\eqref{eq:hamjacreal} is the Bohmian Hamilton-Jacobi equation that contains the local kinetic energy, potential energy, and what is usually known as the \emph{quantum potential energy} 
\begin{align}\label{eq:qpotential}
  Q(x,t) &\equiv \frac{\text{Var}[\text{Re}\,p(x,t)]}{2m} = -\frac{\hbar^2}{2m}\frac{\partial_x^2\sqrt{\rho(x,t)}}{\sqrt{\rho(x,t)}},
\end{align}
which is proportional to the \emph{weak momentum variance} 
\begin{align}\label{eq:wvar}
  \text{Var}[\text{Re}\,p(x,t)] &= \text{Re}\frac{\bra{x}[\op{p}-\text{Re}p(x,t)]^2\ket{\psi}}{\braket{x}{\psi}}, \\
  &= \text{Re}\frac{\bra{x}\op{p}^2\ket{\psi}}{\braket{x}{\psi}} - \left[\text{Re}\frac{\bra{x}\op{p}\ket{\psi}}{\braket{x}{\psi}}\right]^2. \nonumber
\end{align}

The classical particle trajectory limit corresponds to the \emph{eikonal approximation}, or \emph{ray approximation}, (familiar from geometrical optics) where we can neglect rapid spatial changes in the probability density $\rho(x,t)$ that arise from the wave-like quantum interference.  This approximation allows us to neglect the spatial derivatives of $\rho(x,t)$ that appear in the weak momentum variance of Eq.~\eqref{eq:wvar} (or quantum potential of Eq.~\eqref{eq:qpotential}), as well as in the osmotic momentum of Eq.~\eqref{eq:osmom}, implying that the local Bohmian momentum field in Eq.~\eqref{eq:bohmmom} has precise and non-fluctuating values at each definite position.  In this limit, Eq.~\eqref{eq:hamjac} becomes purely real and identified with the usual classical Hamilton-Jacobi equation, while the momentum weak value of Eq.~\eqref{eq:momentum} reduces to the purely real Bohmian momentum and becomes identified with the usual classical momentum field, which describes families of parallel classical particle trajectories.

Evidently, at every stage of the usual quantum-classical correspondence through the Hamilton-Jacobi formalism, weak values directly produce the proper classical mean-field dynamical variables, so are the appropriate conditional estimates akin to the expectation values found in the Ehrenfest theorem. This fundamental role of the weak value corroborates the various interpretations we have encountered in this paper, and makes it clear that weak values are an inextricable feature of the quantum formalism that directly pertains to the classical mean-field limit \cite{Dressel2014b}.

\section{Conclusion}\label{sec:conclusion}
Quantum weak values have endured a controversial history, despite the fact that they are essential to the quantum formalism, with their seeming strangeness being a direct consequence of quantum interference. As estimates of observable averages within a bracketed time-window, or as classical dynamical variables for the mean field evolution, their potential for having anomalous values reflects the nonclassicality of the quantum probabilistic structure, and is equivalent to the need for negative quasiprobabilities. This negativity fundamentally arises as an interference effect, both for the classical mean fields and for manifestly quantum systems that permit discrete detection events and entanglement. No classical model can faithfully reproduce the functional dependence of weak value anomalies without also simulating the corresponding features of quantum mechanics.

\begin{acknowledgments}
The author is grateful for discussions with A. N. Jordan, P. B. Dixon, F. Nori, and E. A. Sete. The research was funded by the Office of the Director of National Intelligence (ODNI), Intelligence Advanced Research Projects Activity (IARPA), through the Army Research Office (ARO) Grant No. W911NF-10-1-0334, and also supported from the ARO MURI Grant No. W911NF-11-1-0268.
\end{acknowledgments}

%

\end{document}